\documentclass[mypaper,7pt,twoside]{CoAst}
\usepackage{epsf,graphicx,fancyhdr,amsmath}
\renewcommand{\chaptermark}[1]{\markboth{#1}{}}

\newcommand{\kopf}{\small\itshape Comm. in Asteroseismology \\ Vol. number, publication date (will be inserted in the production process)}

\newcommand{\Authors}[1]{\begin{center}\normalsize\bf\sf #1 \end{center}}

\renewcommand{\author}[1]{\begin{center}\normalsize\bf\sf #1 \end{center}}
\newcommand{\Address}[1]{\begin{center}\small\sf #1 \end{center}}

\renewenvironment{abstract}{\section*{Abstract}\normalsize\sf}{}
\newcommand{\References}[1]{\begin{flushleft}{\large References\\}\vspace*{2mm}\small #1 \end{flushleft}}

\newcommand{\chapterCoAst}[2]{\chapter[\sf\normalsize #1\\ \footnotesize \hspace*{5mm}by #2 \sf\normalsize][]{#1\\}\rhead[\fancyplain{}{\sf\footnotesize \center{#1}}]{\fancyplain{}{\sffamily\thepage}}\lhead[\fancyplain{\kopf}{\sffamily\thepage}]{\fancyplain{\kopf}{\sf\footnotesize \center{#2}}}}

\newcommand{\figureDSSN}[5]{\begin{figure}[#4]
\centering
\includegraphics*[#5]{#1}
\caption{#2}
\label{#3}
\end{figure}}

\renewcommand{\chaptername}{}

\newcommand{\acknowledgments}[1]{\vspace*{5mm}\noindent  \textbf{Acknowledgments.} #1}

\def\rfr{\smallskip\par\noindent
        \hangindent=7truemm
        \hangafter=1}

\begin{document}
\sf

\chapterCoAst{The frequency spectrum of periodically modulated sinusoidal oscillation}
{B. Szeidl \& J. Jurcsik}
\Authors{B.\,Szeidl \& J., Jurcsik}
\Address{Konkoly Observatory of the Hungarian Academy of Sciences. P.O.~Box~67, H-1525 Budapest, Hungary}

\noindent
\begin{abstract}
The mathematical model of periodically amplitude and phase modulated sinusoidal oscillation is studied, and its Fourier spectrum is given analytically. The Fourier spectrum of the model explains the main features of the frequency spectrum of RR Lyrae stars showing light curve modulation called the Blazhko effect: among others the appearance of multiplets, the rapid decrease of their amplitudes in increasing orders, the asymmetry of the amplitudes of the side frequency pairs, and the possibility of the occurrence of frequency doublets instead of triplets in the spectrum. The good agreement of the results of this mathematical model with observational facts favours those physical models of the Blazhko effect which explain the phenomenon as a modulation of the oscillation with the modulation frequency, $f_m$.
\end{abstract}

\section*{Introduction}
In the past years contradictory descriptions have been published about the frequency spectrum of the RR Lyrae stars with modulated light curves. Analyzing the light curves of RRab stars in the large data bases of MACHO and OGLE (Alcock et al. 2003; Moskalik \& Poretti 2003; Collinge et al. 2006) different classifications of the modulation were introduced depending on the number and the separation of the modulation side-frequencies detected in the spectrum. In variables where an equidistant triplet structure around the fundamental mode frequency was found, no further multiplet structure could be identified in these studies. The observations focused on individual Blazhko stars, however, inevitably proved the presence of higher multiplets (quintuplet, septuplet structures) in the spectrum (Hurta et al. 2008; Jurcsik et al. 2008; Kolenberg et al. 2009, Jurcsik et al. 2009a). Up to now the only mathematical model that aimed to describe the full spectrum with multiplets of Blazhko variables was published by Breger \& Kolenberg (2006). However, no detailed confrontation with the observations of the possible predictions of this model e.g., on the amplitudes of the components of the multiplets has been performed.

As the frequency spectrum implies the basic information (amplitudes and phase angles) on the light curve modulation, its correct knowledge and interpretation is very important for proper understanding of the Blazhko effect. If we find the adequate mathematical model that describes all the features of the frequency spectrum of the well-observed Blazhko stars (frequencies, amplitudes and phase relations), it may provide a starting-point and steady base for theoretical investigations. 

\section*{The frequency spectrum of periodically modulated sinusoidal oscillation}

The amplitude and phase modulated sinusoidal oscillation is given by the formula
\begin{equation}\label{eq:1}
m(t) = a\left[1+b\,\sin \left(\Omega t + \varphi_1\right)\right] \sin \left[\omega t + \varphi_0 + c\,\sin \left(\Omega t + \varphi_2\right)\right],
\end{equation}
\noindent where $\omega=2\pi f_0$, $f_0$ is the fundamental frequency, $\Omega = 2\pi f_{\mathrm m}$, $f_{\mathrm m}$ is the frequency of the modulation, $a$, $b$ and $c=2\pi f_0q$ are the amplitudes of the oscillation, the amplitude  and the phase modulations, respectively. $q$ expresses the amplitude of the phase modulation relative to the fundamental period, $\varphi_0$, $\varphi_1$ and $\varphi_2$ denote the phases of the oscillation, amplitude and phase modulations, respectively. Analytically, there is no limit for the parameters $a$, $b$ and $c(q)$, but because of observational constrains we confine our discussion to the parameter values $0\le b \le 1$ and $|c|\le\pi/2$. Solutions for parameters out of these ranges are irrelevant in asteroseismology. 

We do not consider modulation of the oscillation frequency here as the modulated oscillation of the form of
\begin{equation*}
\begin{split}
m(t) = a\,\sin[\omega  (1+d \sin(\Omega +\varphi_3)) t  + \varphi_0]= \\
a\,\sin[\omega t + \varphi_0 + d \omega t \sin(\Omega t+\varphi_3)   ]
\end{split}
\end{equation*}
corresponds to a phase modulation with variable modulation amplitude ($c=d \omega t$) and time dependent, unstable frequency spectrum.

By suitable choice of the starting epoch, without any restriction of the general validity, $\varphi_0=0$ and $\varphi_2=0$ can be attained. If the initial epoch corresponds to the timing of the mid of rising branch of the phase modulation at the moment of the mid of the descending branch of the oscillation (taking into account the reverse direction of the magnitude scale) both $\varphi_0=0$ and $\varphi_2=0$ fulfill. Since $f_0$ and $f_m$ can be regarded as rational numbers as the accuracy of their numerical value is limited by the observations, such an epoch should exist. We note here that with cosine representation the choice of the appropriate initial epoch would be more natural, it would correspond to the maxima of the oscillation and the phase modulation.
 
Denote now $\Phi=\varphi_1-\varphi_2$ the epoch independent phase difference between the amplitude and phase modulations. Then the time history of the modulated oscillation is described as follows
\begin{equation}\label{eq:2}
m(t) = a\left[1+b\,\sin\left(\Omega t + \Phi\right)\right] \sin \left(\omega t + c\,\sin \Omega t\right).
\end{equation}

Taking the simple trigonometric addition formula of $\sin(\alpha + \beta)$ and the Taylor-series of $\sin x$  and $\cos x$ into account, Eq.~\ref{eq:2} has the form
\begin{equation}\label{eq:3}
  \begin{split}
    m(t)& = a\left(1+b\,\sin \Phi \cos \Omega t + b\,\cos \Phi \sin \Omega t\right)\cdot{}\\
&\left[\sin\omega t\sum_{n=0}^\infty(-1)^n\frac{c^{2n}}{(2n)!}\sin^{2n}\Omega t+\cos\omega t\sum_{n=0}^\infty(-1)^n\frac{c^{2n+1}}{(2n+1)!}\sin^{2n+1}\Omega t\right]
  \end{split}
\end{equation}
Substituting  the well-known power-reduction formulae Eqs.~\ref{eq:4} and \ref{eq:5} into Eq.~\ref{eq:3} with $\alpha = \Omega t$
\begin{equation}\label{eq:4}
  \sin^{2n}\alpha=\frac{1}{2^{2n-1}}\left[-\frac{1}{2} \binom{2n}{n}+\sum_{k=0}^n (-1)^k\binom{2n}{n-k}\cos2k\alpha\right]
\end{equation}
\noindent and
\begin{equation}\label{eq:5}
  \sin^{2n+1}\alpha=\frac{1}{2^{2n}}\sum_{k=0}^n (-1)^k\binom{2n+1}{n-k}\sin(2k+1)\alpha
\end{equation}
\noindent and then applying the simple trigonometric formulae
\begin{equation}\label{eq:6}
  \begin{split}
    \sin \alpha \sin \beta &= \frac{1}{2}\left[\cos\left(\alpha-\beta\right)-\cos\left(\alpha+\beta\right)\right] \\
    \sin \alpha \cos \beta &= \frac{1}{2}\left[\sin\left(\alpha-\beta\right)+\sin\left(\alpha+\beta\right)\right] \\
    \cos \alpha \cos \beta &= \frac{1}{2}\left[\cos\left(\alpha-\beta\right)+\cos\left(\alpha+\beta\right)\right]
  \end{split}
\end{equation}
\noindent where $\alpha = \omega t$, $\beta = i\Omega t, i = 1,2,3, \dots$, Eq.~\ref{eq:3} will have the form
\begin{equation}\label{eq:7}
  \begin{split}
  m(t)  = x_0 \sin\omega t + z_0 \cos \omega t& + \sum_{i=1}^\infty x_i \sin(\omega + i \Omega)t+\sum_{i=1}^\infty y_i \sin (\omega - i\Omega)t \\
    &+\sum_{i=1}^\infty z_i \cos(\omega+i\Omega)t + \sum_{i=1}^\infty w_i \cos (\omega - i\Omega)t
  \end{split}
\end{equation}
\noindent where $x_i$, $y_i$, $z_i$ and $w_i$ ($i=1,2,3,\dots$) coefficients depend only~on~$a$,~$b$,~$c$~and~$\Phi$.

Applying the known relations Eqs.~\ref{eq:8} and \ref{eq:9} to Eq.~\ref{eq:7}:
\begin{equation}\label{eq:8}
  \cos\alpha = sin\left(\alpha+\frac{\pi}{2}\right),\ \ \  \alpha=\left(\omega \pm i\Omega\right)t\ \ \ i=0, 1, 2, \dots
\end{equation}
\noindent and
\begin{equation}\label{eq:9}
  X_1 \sin\nu t + X_2\sin\left(\nu t + \frac{\pi}{4}\right) = X\sin\left(\nu t + \varphi\right),\ \ \  \nu = \omega \pm i\Omega\ \ \ (i = 0, 1, 2, \dots)
\end{equation}
\noindent where $X^2 = X_1^2 + X_2^2$ and $\tan\varphi = {X_1}/{X_2}$ we arrive at the Fourier spectrum of the modulated sinusoidal oscillation:
\begin{equation}\label{eq:10}
  \begin{split}
    m(t) &= A_0\sin(\omega t + \chi_0) + \sum_{i=1}^\infty A_i^+\sin\left[\left( \omega + i\Omega\right)t+\chi_i^+\right] \\
    &\quad {} + \sum_{i=1}^\infty A_i^-\sin\left[\left( \omega - i\Omega\right)t+\chi_i^-\right]
  \end{split}
\end{equation}

By the outlined procedure we can derive the $A_0$, $A_i^+$ and $A_i^-$ coefficients (amplitudes) as well as their phase angles, $\chi_0$, $\chi_i^+$ and $\chi_i^-$. 

As an example we derive the amplitudes and phases of the triplet. In this case only the $k=0, 1$ terms in Eq.~\ref{eq:4} and the $k=0$ term in Eq.~\ref{eq:5} should be considered and substituted into Eq.~\ref{eq:3} for each $n$ ($n=0,1,2, \dots $). All the other terms contribute to the higher members of the multiplets. In this case taking Eqs.~\ref{eq:6} into account, Eq.~\ref{eq:3} takes the form:
\begin{equation}\label{eq:11}
  \begin{split}
    m(t) &= a\left(1+b\sin\Phi\cos\Omega t + b\cos\Phi\sin\Omega t\right)\cdot{}\\
    &\cdot \left\{\sin\omega t \sum_{n=0}^\infty (-1)^n \frac{c^{2n}}{2^{2n-1}(2n)!}\left[\binom{2n}{n} - \binom{2n}{n-1} \cos 2\Omega t - \frac{1}{2}\binom{2n}{n}\right]\right. \\
    &+\left.\cos\omega t \sum_{n=0}^\infty (-1)^n \frac{c^{2n+1}}{2^{2n}(2n+1)!} \binom{2n+1}{n}\sin\Omega t\right\} + g
  \end{split}
\end{equation}
\noindent where $g$ is a function of $\omega \pm k\Omega, k\geq 2$. 

Let the convergent series
\begin{equation}\label{eq:12}
  S=\sum_{n=0}^\infty (-1)^n \left(\frac{c^n}{2^nn!}\right)^2 = 1 - \frac{c^2}{4}+\frac{c^4}{64} - \frac{c^6}{2304} + \frac{c^8}{147456} - \frac{c^{10}}{14745600} + \dots 
\end{equation}
\noindent and
\begin{equation}\label{eq:13}
  S_1=\sum_{n=0}^\infty (-1)^n \left(\frac{c^n}{2^nn!}\right)^2 \frac{n}{n+1} = - \frac{c^2}{8}+\frac{c^4}{96} - \frac{c^6}{3072} + \frac{c^8}{184320} - \frac{c^{10}}{17694720} + \dots
\end{equation}
\noindent Since 
\begin{equation*}
\binom{2n}{n}=\frac{(2n)!}{(n!)^2};\,\,\,\,\, \binom{2n}{n-1}=\frac{(2n)!}{(n!)^2}\frac{n}{n+1};\,\,\,\,\, \binom{2n+1}{n}=\frac{(2n+1)!}{(n!)^2}\frac{1}{n+1}
\end{equation*}
\noindent Eq.~\ref{eq:11} takes the form:
\begin{equation*}
  m(t) = a\left(1+b\sin\Phi\cos\Omega t + b\cos\Phi\sin\Omega t\right)\,\cdot\nonumber
\end{equation*}
\begin{equation}\label{eq:14}
  \cdot\,\left[S\sin\omega t - 2 S_1\sin\omega t \cos 2\Omega t + (S-S_1)c \cos \omega t \sin \Omega t \right] + g
\end{equation}

If we execute the multiplications in Eq.~\ref{eq:14} taking into account Eqs.~\ref{eq:6}, and disregard all the terms with $\omega\pm k\Omega, k\geq 2$, we obtain the following equation that describes the triplet structure:
\begin{equation}\label{eq:15}
  \begin{split}
    m(t)_\mathrm{triplet} &= a\left\{\left[ S \sin\omega t + \frac{1}{2}(S-S_1)b c \cos\Phi\cos\omega t \right]\right. \\
    &\quad{}+\left[\frac{1}{2} S b\sin\Phi - \frac{1}{2}S_1 b \sin\Phi - \frac{1}{2}(S-S_1) c\right]\sin(\omega-\Omega)t \\
    &\quad{}+\left[\frac{1}{2} S b\cos\Phi + \frac{1}{2}S_1 b \cos\Phi \right]\cos(\omega-\Omega)t \\
    &\quad{}+\left[\frac{1}{2} S b\sin\Phi - \frac{1}{2}S_1 b \sin\Phi + \frac{1}{2}(S-S_1) c\right]\sin(\omega+\Omega)t \\
    &\quad{}+\left[\left.-\frac{1}{2} S b\cos\Phi + \frac{1}{2}S_1 b \cos\Phi \right]\cos(\omega+\Omega)t\right\}
  \end{split}
\end{equation}

Application of Eqs.~\ref{eq:8} and \ref{eq:9} to Eq.~\ref{eq:15} yields the squares of the amplitudes as well as the phases of the triplet:
\begin{equation}\label{eq:16}
  A_0^2 = a^2[S^2 + \frac{1}{4}(S-S_1)^2b^2c^2\cos^2\Phi]
\end{equation}
\begin{equation}\label{eq:17}
  \begin{split}
    (A_1^+)^2 &= a^2 \left[ \frac{1}{4}(S-S_1)^2(b^2+c^2+2bc\sin\Phi)+SS_1b^2\cos^2\Phi\right]\\
    (A_1^-)^2 &= a^2 \left[ \frac{1}{4}(S-S_1)^2(b^2+c^2-2bc\sin\Phi)+SS_1b^2\cos^2\Phi\right]
  \end{split}
\end{equation}
\begin{equation}\label{eq:18}
  \tan\chi_0 = \frac{1}{2} \frac{S-S_1}{S}bc\cos \Phi
\end{equation}
\begin{equation}\label{eq:19}
  \begin{split}
    \tan\chi_1^+ &= -\frac{(S+S_1)b\cos\Phi}{(S-S_1)(b\sin\Phi+c)}\\
    \tan\chi_1^- &= -\frac{(S+S_1)b\cos\Phi}{(S-S_1)(b\sin\Phi-c)}
  \end{split}
\end{equation}

Note that a remarkable relation exists for the power difference of the side frequencies:
\begin{equation}\label{eq:20}
(A_1^+)^2-(A_1^-)^2=(S-S_1)^2 a^2 b c \sin\Phi
\end{equation}
\noindent Eq.~\ref{eq:20} shows that the asymmetry of the side frequency amplitudes depends basically on the phase difference between the amplitude and phase modulation components.

The amplitudes and phases of the subsequent frequencies of the multiplets can be derived in the same way. E.g., the final result for the side frequencies of the quintuplet ($k=2$) is as follows. Let the convergent series
\begin{equation*}
  \begin{split}
    S_2 &= \sum_{n=0}^\infty (-1)^n \left(\frac{c^n}{2^nn!}\right)^2\frac{n}{(n+1)(n+2)} \\
    & = - \frac{c^2}{24}+\frac{c^4}{384} - \frac{c^6}{15360} + \frac{c^8}{1105920} - \frac{c^{10}}{123863040} + \dots
  \end{split}
\end{equation*}
\noindent then
\begin{equation*}
  \begin{split}
    (A_2^+)^2 &= a^2 \left[ S_1^2 - \frac{1}{2}(S-S_1-S_2) bc\sin\Phi \right.\\
    &\quad\left.{}+ \frac{1}{16}(S-S_1-S_2)^2b^2c^2 + \frac{1}{4}S_2(S-S_1)b^2c^2\cos^2\Phi\right]
  \end{split}
\end{equation*}
\noindent and
\begin{equation*}
  \begin{split}
    (A_2^-)^2 &= a^2 \left[ S_1^2 + \frac{1}{2}(S-S_1-S_2) bc\sin\Phi \right.\\
    &\quad\left.{}+ \frac{1}{16}(S-S_1-S_2)^2b^2c^2 + \frac{1}{4}S_2(S-S_1)b^2c^2\cos^2\Phi\right]
  \end{split}
\end{equation*}
\noindent while
\begin{equation*}
  \tan\chi_2^+ = \frac{(S-S_1+S_2)bc\cos\Phi}{4S_1-(S-S_1-S_2)bc\sin\Phi}
\end{equation*}
\begin{equation*}
 \tan\chi_2^- = \frac{(S-S_1+S_2)bc\cos\Phi}{4S_1+(S-S_1-S_2)bc\sin\Phi}.
\end{equation*}

Note again that the power difference of the frequency pair depends simply on $\sin \Phi$:
\begin{equation*}
  (A_2^+)^2 -(A_2^-)^2 =-(S-S_1-S_2) a^2 b c \sin \Phi .
\end{equation*}

\section*{Discussion}

In realistic cases, in Blazhko stars the harmonics of the fundamental frequency are also present in the oscillation and appear in the frequency spectrum. So the time history of the periodically modulated oscillation is described by the equation 

\begin{equation}\label{eq:21}
m(t) = \sum_{i=1}^\infty a_i\left[1+b_i\,\sin\left(\Omega t + \Phi_{i}\right)\right] \sin \left[i\omega t + c_i\,\sin \Omega t\right].
\end{equation}

In the previous section we confined ourselves to the fundamental frequency, but the same deduction is valid and true for the harmonics too, simply $\omega$ should be replaced by $i\omega$. Therefore, we conclude that the frequency spectrum of any periodically modulated periodic oscillation i.e. the light curve of Blazhko stars is the infinite series of the multiplets of the fundamental frequency and its harmonics:

\begin{equation}\label{eq:22}
  \begin{split}
    m(t) &= \sum_{i=1}^\infty A_{i0}\sin(i\omega t + \chi_{i0}) + \sum_{i=1}^\infty \sum_{j=1}^\infty A_{ij}^+\sin\left[\left( i\omega + j\Omega\right)t+\chi_{ij}^+\right] \\
    &\quad {} + \sum_{i=1}^\infty \sum_{j=1}^\infty A_{ij}^-\sin\left[\left( i\omega - j\Omega\right)t+\chi_{ij}^-\right]
  \end{split}
\end{equation}

We also note here that, if the sums in Eq.~\ref{eq:22} start from $i=0$, then this description gives a natural ground to the appearance of $A_{00}$, and frequencies at $jf_m$, which correspond to an arbitrary zero point of the scale, and the modulation frequency and its harmonics, respectively. Although the appearance of $f_m$ (and perhaps $2f_m$) is hardly (if at all) discernible in most of the observed frequency spectrum of Blazhko stars, in accurate, extended data the modulation frequency can be always detected with very small amplitude most probably because the star's physical parameters (e.g. mean luminosity, temperature) change very weakly during the Blazhko cycle (Jurcsik et al. 2009a, 2009b).

\figureDSSN{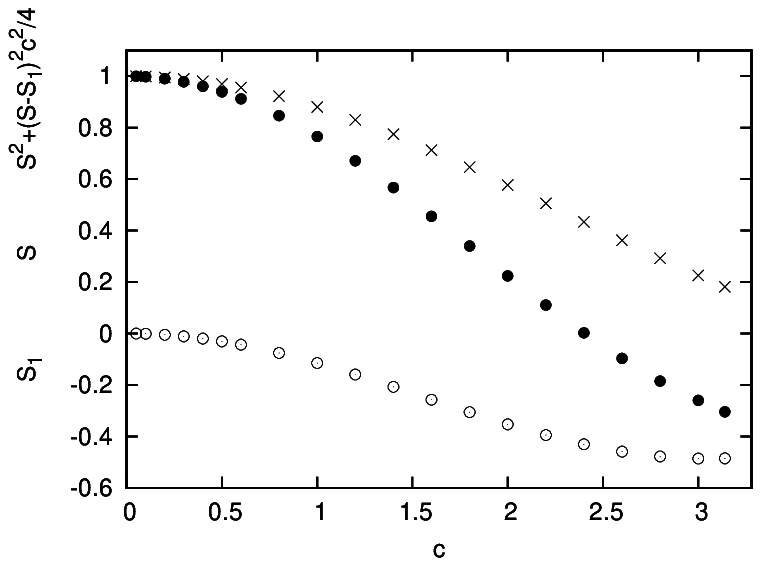}{Variations of the $S$ (dots), $S_1$ (circles) sums and their combination term (crosses) which define the amplitude decrease of the $f_0$ frequency component with increasing phase modulation amplitude, $c$.  }{sgen}{!ht}{clip,angle=0,width=105mm}

A priori there is no definite connection between the parameters of the modulation. The rough constancy of the phase differences and the systematic behaviour of the amplitudes of the triplet components with increasing orders $i$  (see Fig. 10  in Jurcsik et al. 2009a) suggests, however, the existence of some connection among them. If we suppose that $b$, $c$  and $\Phi$ are unique parameters of the modulation in the different harmonic orders then $A^{+/-}_{ij}$ and  $\chi_{ij}^{+/-}$ would depend only on $a_i$. In this case  the side frequencies of the different order pulsation components would naturally have some common properties.

In all probability the amplitudes of the $if_0$ frequencies in the spectrum of modulated light curves run more or less similarly to that of the single periodic RR Lyraes. It should, however, be noted that according to Eq.~\ref{eq:16}, $a^2[S^2 + (S-S_1)^2c^2/4]$ is an upper limit ($b=1, \Phi=0$) for the amplitude of $f_0$, i.e., $A_{0}/a=1$ only if the modulation is pure amplitude modulation ($c=0$). Any phase modulation component lowers the amplitude of the $f_0$ frequency.  Fig.~\ref{sgen} shows the dependence of $S, S_1$ and the upper limit of the amplitude reduction factor $(S^2 + (S-S_1)^2c^2/4$) on $c$, using $f_0=2.1925$ c/d oscillation frequency, corresponding to the pulsation frequency of DM Cyg. The phase modulation lowers the amplitudes of the harmonic components, too,  ($A_{i0}/a_i<1$ if $c_i>0$) as E.q.~\ref{eq:16} holds for the frequencies $\omega=i\omega$, too.

As mentioned above, the accurate and well-distributed (over both frequencies $f_0$ and $f_m$) observations of Blazhko stars clearly show the multiplet structure around the fundamental mode frequency and its harmonics in their Fourier spectrum (Jurcsik et al. 2008, Kolenberg et al. 2009). The question arises why the triplet is a striking feature while the higher multiplets are hardly perceptible in the frequency spectrum of a Blazhko star.

\figureDSSN{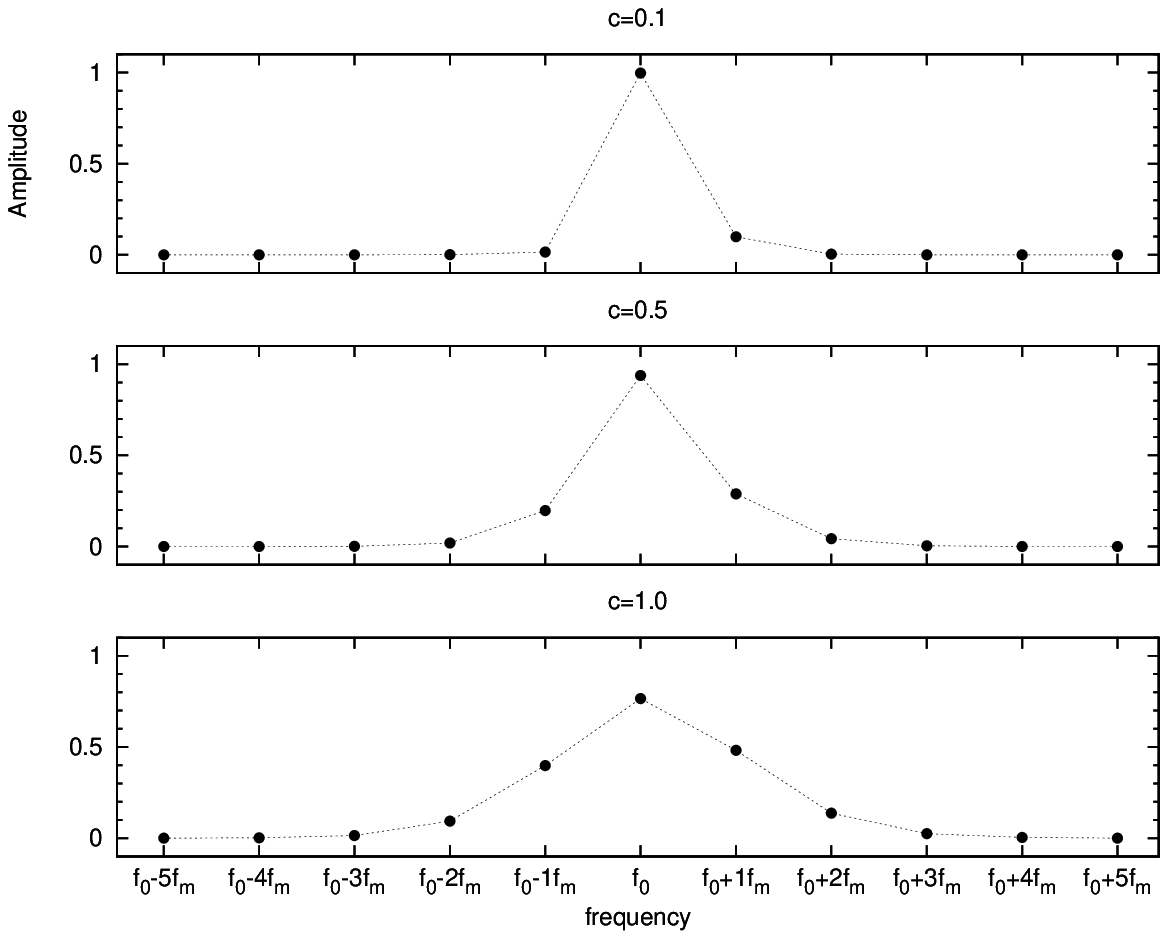}{Amplitude decrease of the multiplet components in the different orders of the modulation. Artificial datasets were generated for different values of $c$ (0.1, 0.5, 1.0) according to Eq. 3 using $a=1$,  $b=0.1$ and $\Phi=72^\circ$ values and $f_0, f_m$ frequencies of DM Cyg. The phase modulation with $c=1$ amplitude corresponds to full amplitude of the phase modulation slightly less than 1/3 of the main period which is substantially larger than the amplitude of the phase modulation observed in Blazhko stars, which is typically about $0.1-0.2$ pulsation phase. Even in the $c=1$ simulations the amplitudes of the higher order modulation components ($f_0\pm kf_m$, $k\ge 3$) are neglectably small.
}{param1}{!ht}{clip,angle=0,width=125mm}

\begin{table}
\begin{center}
\caption{Amplitudes of the frequency multiplets appearing in the Fourier spectrum of amplitude and phase modulated sinusoidal signal according to Eq.~\ref{eq:2} using $a=1, b=0.1, c=0.5$ parameters for different $\Phi$ values.}\label{table}
\begin{tabular}{ccccccc}
\hline
 $\Phi$&$f_0-5f_m$&  $f_0-4f_m$&  $f_0-3f_m$&  $f_0-2f_m$&  $f_0-f_m$&  $f_0$ \\
\hline
$36^\circ$&0.000007&0.000134&0.002067&0.025337&0.216919&0.938674\\
$72^\circ$&0.000003&0.000055&0.001197&0.019320&0.196687&0.938500\\
$90^\circ$&0.000000&0.000032&0.001025&0.018363&0.193815&0.938470\\
\\
\hline
 $\Phi$&$f_0+5f_m$&  $f_0+4f_m$&  $f_0+3f_m$&  $f_0+2f_m$&  $f_0+f_m$&  \\
\hline
$36^\circ$&0.000014&0.000258&0.003680&0.039023&0.273228&\\
$72^\circ$&0.000016&0.000286&0.004054&0.042408&0.288692&\\
$90^\circ$&0.000016&0.000289&0.004102&0.042845&0.290722&\\
\hline
\end{tabular}
\end{center}
\end{table}

\figureDSSN{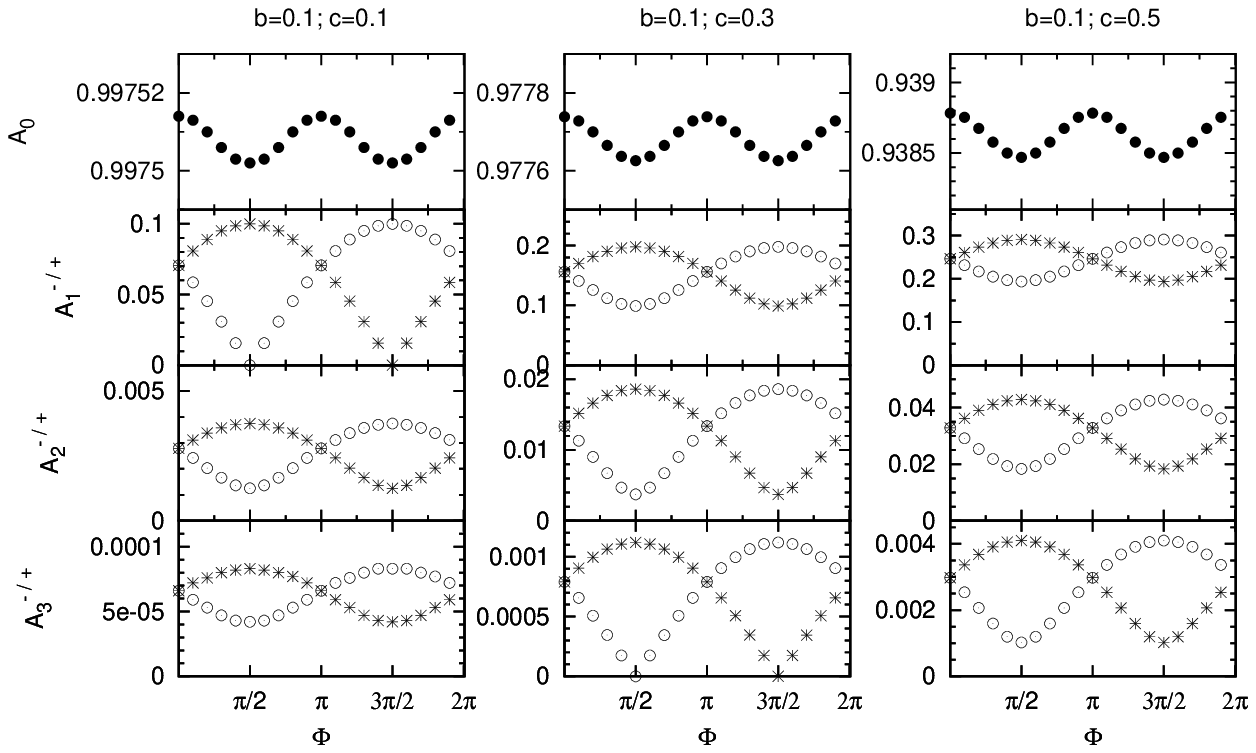}{Amplitude variations of the multiplet frequency components of periodically amplitude and phase modulated sinusoidal signal as a function of $\Phi$, the phase difference between the amplitude and phase modulation components. $A_{0}, A^{+/-}_{1}, A^{+/-}_{2}, A^{+/-}_{3}$ denote the amplitudes of the $f_0, f_0\pm f_m,  f_0\pm 2f_m$, and $f_0\pm 3f_m$ frequency components. The three panels show the results for different combinations of $b$ and $c$, the amplitudes of the amplitude and phase modulations, respectively. Artificial datasets were generated using Eq.~\ref{eq:2} with  $a=1$ and  $f_0 =  2.3817$, $f_m= 0.06043$ c/d values corresponding to the pulsation and modulation frequencies of DM Cyg, and Fourier analysed in order to determine the amplitude values. }{param}{!ht}{clip,angle=0,width=115mm}

Although the formulae of the amplitudes of the multiplets can be exactly derived as it was shown in the previous section, they are fairly complicated expressions (besides the outlined procedure is too laborious). Their behaviour can be, however, easily studied through the Fourier analysis of synthetic light curves. As examples, Figs.~\ref{param1},~\ref{param} and Table~\ref{table} show the amplitudes of subsequent multiplets of the frequency spectra of Eq.~\ref{eq:2} at different choice of the modulation parameters. The striking feature is that the amplitudes of the subsequent frequencies on both sides of the fundamental frequency approach zero rapidly. The degree of the decrease depends on the parameter of the phase modulation $c$, the weaker the phase modulation, the speedier the amplitude decrease is. Even if the phase modulation has a high value, $c=1$, the amplitude of the  $f_0\pm5f_m$ frequencies are more than three orders of magnitude less than the amplitude of $f_0$ (see Fig~\ref{param1}). For a more realistic case e.g., with $c=0.5$ the amplitude difference between the first and fifth order modulation components is larger than four orders of magnitude as shown in Table~\ref{table}. It is now clear that only the observational accuracy sets limit to the perception of higher multiplets in the frequency spectrum.

One further important implication of our results is that Eq.~\ref{eq:20} proves that the power difference of the side frequencies in the triplet is the physically meaningful quantity to measure the asymmetry of the triplet instead of their amplitude ratios. 

Our calculations have further serious and amazing consequences on the symmetry and asymmetry of the triplet components. It results also from Eq.~\ref{eq:20} that the triplet will only be symmetrical if any of the quantities $b$, $c$ or $\Phi$ equals to zero, or $\Phi=\pi$. If the quantities $b$, $c$ and $\Phi$ have appropriate numerical values, one of the amplitudes of the triplet side frequencies $A^+_1$ or $A^-_1$ may become zero. The only solution of the $(A^+_1)^2=0$ or $(A^-_1)^2=0$ second order equations (Eqs.~\ref{eq:17}) is $b=c$ and $\Phi=3\pi/2$ or $\Phi=\pi/2$. In this case the spectrum of the modulated oscillation is doublet instead of triplet. As an instructive example Fig.~\ref{param} shows the amplitude changes of the frequencies as a function of $\Phi$ for $a=1, b=0.1$ and $c=0.1, 0.5, 1.0 $. In Blazhko stars, usually there is a phase difference between the amplitude and phase modulation components (see e.g., Fig 8 in Jurcsik et al 2008, and Fig 3 in Jurcsik et al 2009a) and this explains why asymmetric triplets are detected in most of the cases.

Concerning the phases, from Eq.~\ref{eq:18} it follows that if either $b=0$ or $c=0$  the phase of the oscillation is $0$ or $\pi$ ($\tan \chi_0 =0$) if the initial epoch is chosen as described in the beginning of Sect. 1. When the modulation is pure phase modulation, then $\tan \chi_1^- = \tan \chi_1^+ =0$, i.e.,  $\chi_1^-$, and $\chi_1^+$ are $0$ or $\pi$. When the modulation is pure amplitude modulation, then Eq. 1 can be easily solved with appropriate choices of the initial epoch. If the initial epoch is chosen to fulfill $\varphi_0=0$ and $\varphi_1=0$, i.e., it corresponds to the timing of the mid of rising branch of the amplitude modulation at the moment of the mid of the descending branch of the oscillation, the solution of Eq. 1 shows  that  $\chi_1^-=\pi/2$ and $\chi_1^+=-\pi/2$. Alternatively, if the initial epoch corresponds to the maximum phase of the amplitude modulation i.e., $\varphi_1=\pi/2$, then $\chi_1^-= \chi_1^+ =\chi_0 $ fulfill.

Eq.~\ref{eq:20} provides another interesting result. If $0 <  \Phi  <  \pi$, then $A^+_1 > A^-_1$ and the plot that shows the  amplitude of the light variation vs. phase shift of maximum light during the Blazhko cycle has anticlockwise progression, while in the case of $-\pi<\Phi < 0$,  $A^+_1<A^-_1$, the progression is the opposite. In reality, of course, the situation is somewhat more complex because the modulations of the harmonic components may modify the picture.  

\section{Conclusions}

The model of periodically modulated harmonic oscillation properly explains the main features of the frequency spectrum of Blazhko RR Lyrae stars. It predicts the infinite multiplet systems around the fundamental frequency and its harmonics, accounts for the often detected asymmetry of the amplitudes of the side frequency pairs in the triplets (quintuplets) and for the rapid decrease of the amplitudes of the multiplets with increasing orders.

Our results also indicate that the suggested classification schemes of Blazhko stars based on their frequency spectrum (Alcock et al. 2003,  Moskalik and Poretti 2003) are dubious. For example, the frequency doublets, which are generally interpreted with a nonradial frequency component close to the radial frequency, can naturally take its origin from amplitude and phase modulations of a single oscillation. Our analysis also show that the asymmetry of the triplets are a natural consequence of the mixture of amplitude and phase modulations. We thus conclude that the occurrence of multiplets does not necessarily imply the presence of more than a single oscillation, and prefers those physical models which connect the phenomenon to one modulation frequency, $f_m$. Recently, Stothers (2006) proposed such a model for Blazhko RR Lyrae stars. In his model the cyclic changes in the strength of the envelope convection gives rise to the modulation of the periodic oscillation of these stars.

In the near future we plan further investigations to exploit the potentials of this model and to find connections between the amplitudes and phase angles in the frequency spectrum. We also plan to carry out a more profound comparison with observational facts.

\acknowledgments{We thank the referee, Katrien Kolenberg for her helpful comments. We are grateful to \'Ad\'am S\'odor for his help in preparing the manuscript. The financial support of OTKA grant T-068626 is acknowledged. } 

\References{
\rfr Alcock, C., Alves, D. R., Becker, A., et al. 2003, ApJ, 598, 597
\rfr Breger, M., \& Kolenberg, K. 2006, A\&A 460, 167
\rfr Collinge, M. J., Sumi, T., \& Fabrycky, D. 2006, ApJ, 651, 197
\rfr Hurta, Zs., Jurcsik, J., Szeidl, B., S\'odor. \'A. 2008, AJ, 135, 957
\rfr Jurcsik, J., Hurta, Zs., S\'odor, \'A., et al. 2009a, MNRAS accepted
\rfr Jurcsik, J., S\'odor, \'A., Hurta, Zs.,  et al. 2008, MNRAS, 391, 164
\rfr Jurcsik, J., S\'odor, \'A., Szeidl, B., et al. 2009b, MNRAS, 393, 1553
\rfr Kolenberg, K., Guggenberger, T., Medupe, T., et al. 2009, MNRAS, 396, 263
\rfr Moskalik, P., \& Poretti, E. 2003, A\&A, 398, 213
\rfr Stothers, R. 2006, ApJ, 652, 643
}

\end{document}